\documentclass{mn2e}
\usepackage{psfig}

\def\ltsima{$\; \buildrel < \over \sim \;$}
\def\lsim{\lower.5ex\hbox{\ltsima}}
\def\gtsima{$\; \buildrel > \over \sim \;$}
\def\gsim{\lower.5ex\hbox{\gtsima}}

\begin{document}
\title[Non-LTE dust nucleation]{Non-LTE dust nucleation in
sub-saturated vapors} \author[Lazzati] {Davide Lazzati \\ JILA,
University of Colorado, 440 UCB, Boulder, CO 80309-0440, USA}

\maketitle

\begin{abstract}
We use the kinetic theory of nucleation to explore the properties of
dust nucleation in sub-saturated vapors. Due to radiation losses, the
sub-critical clusters have a smaller temperature compared to their
vapor. This alters the dynamical balance between attachment and
detachment of monomers, allowing for stable nucleation of grains in
vapors that are sub-saturated for their temperature. We find this
effect particularly important at low densities and in the absence of a
strong background radiation field. We find new conditions for stable
nucleation in the $n-T$ phase diagram. The nucleation in the non-LTE
regions is likely to be at much slower rate than in the
super-saturated vapors. We evaluate the nucleation rate, warning the
reader that it does depend on poorly substantiated properties of the
macro-molecules assumed in the computation. On the other hand, the
conditions for nucleation depend only on the properties of the large
stable grains and are more robust. We finally point out that this
mechanism may be relevant in the early universe as an initial dust
pollution mechanism, since once the interstellar medium is polluted
with dust, mantle growth is likely to be dominant over non-LTE
nucleation in the diffuse medium.
\end{abstract}


\section{Introduction}

Dust particles are one of the fundamental components of the
interstellar medium (ISM) and an ever-present worry for observers due
to their opacity at optical and UV wavelengths (Cardelli, Clayton \&
Mathis 1989). The ISM of the Milky Way is polluted by a mixture of
grains made of a variety of materials, likely dominated by
carbonaceous grains, silicates, and small PAHs particles (Mathis,
Rumpl \& Nordsieck 1977; Weingartner \& Draine 2001). The dust
properties are supposed to be the result of dust formation in the
outflows of evolved stars (e.g. Salpeter1977; Stein \& Soifer 1983;
Mathis 1990; Whittet 1992; Draine 2003, and references therein) and
subsequent evolution, and eventual dissolution, in the ISM, mainly as
the effect of shock waves that destroy the grains through sputtering
(Draine 1989; McKee 1989; Edmunds 2001). Alternative dust production
sites are supernova explosions (Kozasa, Hasegawa \& Nomoto 1989, 1991;
Todini \& Ferrara 2001; Nozawa et al. 2003; Schneider, Ferrara \&
Salvaterra 2004; Bianchi \& Schneider 2007) and quasar outflows
(Elvis, Marengo \& Karovska, 2002).

The theory of dust nucleation in astrophysics is heavily influenced by
the theory of the nucleation of phase transitions in super-saturated
vapors (Becker \& Doring 1935; Feder et al. 1966; Abraham 1974). The
theory had mild success in reproducing nucleation rates, but is still
controversial in many aspects, especially because it extrapolates the
properties of macroscopic bodies to clusters of few molecules and
because it extends the thermodynamic approach to systems with a
handful of particles. To add to these problems, astrophysical dust
nucleation requires chemical reactions, since grains of materials that
do not have a vapor state do nucleate (think, for example, to the
nucleation of olivines from silicon oxides and metals; Draine 1979;
Gail \& Sedlmayr 1986). An alternative approach is the so-called
kinetic theory, which describes nucleation as the result of attachment
and detachment of monomers from a seed cluster of $n$ particles
(atoms, molecules or radicals; Nowakowski \& Ruckenstein 1991ab).

Both the thermodynamic and the kinetic nucleation theory have been
developed in conditions of {\it true equilibrium}, i.e., when the two
phases have the same temperature. In the astrophysical scenario,
however, the temperature of the dust grains can be sensibly lower than
the temperature of the gas in which they are embedded due to efficient
radiation cooling (e.g., Draine 1981). This would seem to be
irrelevant to nucleation theory, since a vapor needs to be already
nucleated in order to have grains that can be colder than the gas
phase. Even a sub-saturated vapor, however, has a large number
of unstable clusters that form by random association of monomers (and
rapidly evaporate). In this paper we study the effect of cooling of
these proto-clusters in a sub-saturated vapor, and the effect this has
on the balance between attachment and detachment of monomers. Using
the kinetic theory of nucleation, we find that even largely
sub-saturated vapors can nucleate, provided they are not immersed in a
strong radiation field.  We compute, albeit under some
controversial assumptions, the non-LTE nucleation rate. We show that,
even though it is not as large as in super-saturated vapors, it can
produce dust grains at a rate that can reproduce the average dust
grain density in the Milky Way over a timescale of several million
years. In addition we show that non-LTE effects can increase the
nucleation rate in super-saturated vapors. Non-LTE nucleation could
therefore provide a slow channel for dust formation, in which dust is
built over a relative long time in a slowly evolving region. Such an
example could be the outflow from AGN nuclei (Elvis et al. 2002). Such
evolution is different from the one envisaged in the classical dust
factories -- AGB star atmospheres and supernov\ae\ -- where dust
nucleation is rapid but short lived since the favorable conditions are
rapidly lost.

This paper is organized as follows: in \S~2 we briefly review the
classical kinetic theory of nucleation; in \S~3 we compute the dust
grain temperature and in \S~4 we compute the new conditions for
nucleation. In \S~5 we consider the nucleation rate and discuss our
results in \S~6.

\section{Classical nucleation}

In this section we review some basic concepts of the theory of
nucleation at densities and temperatures typical of laboratory
experiments (in conditions of ``true equilibrium''). Let us first
consider a flat surface separating the vapor phase of a certain
material with its condensed phase (either liquid or solid). The
equilibrium condition between the two phases can be described in
either thermodynamic or kinetic terms.

In thermodynamic terms, equilibrium implies that the temperature of
the two phases are equal and that the chemical potential of the
molecules in the two phases are equal (e.g., Vehkam\"aki 2006). In
this paper we adopt the kinetic approach, since we will investigate
nucleation out of thermal equilibrium. In the kinetic representation
the equilibrium is dynamical and is obtained by equating the rate of
vapor phase molecules that become attached to the condensed phase to
the rate of molecules that return from the condensed phase to the
vapor. The first rate is relatively easy to quantify, as long as it
can be assumed that the vapor molecule velocity distribution is
Maxwellian\footnote{We concentrate here on homogeneous nucleation in
vapors with only inert species and the compound of interest.}:
\begin{equation}
\frac{dn_{\rm{in}}}{dt\;dA} = k_s n_X\sqrt{
\frac{kT}{2\pi m_X}}
\label{eq:attach}
\end{equation}
where $n_{X}$ is the density of the molecules (or atoms) of the
compound of interest in the vapor phase, $k$ is the Boltzmann
constant, $T$ is the temperature and $m_X$ the molecular mass. The
parameter $k_s\le1$ is the sticking coefficient and represents the
probability that the incoming monomer remains attached to the
condensed phase rather than bounce and return to the vapor. There is
no reason for $k_s$ to be constant and it does indeed depend on the
temperature (see, e.g., Batista et al. 2005). For the sake of
simplicity, we will consider here $k_s$ to be a constant.

\begin{figure}
\psfig{file=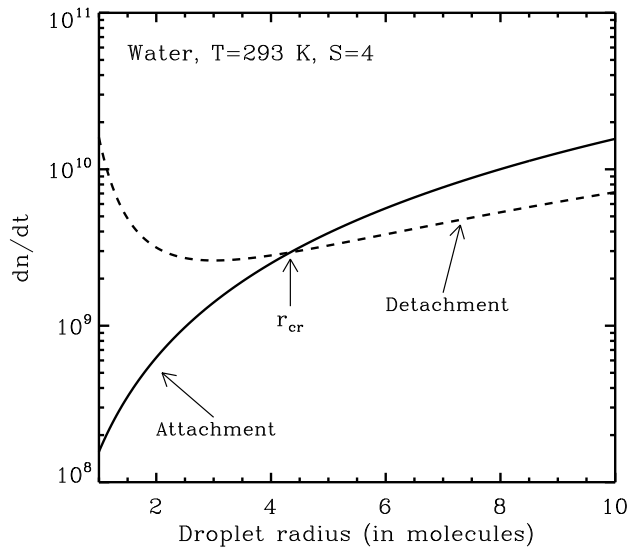,width=\columnwidth}
\caption{{The attachment and detachment of monomers from a water
droplet at $T=293$~K and saturation $S=4$. The solid curve shows the
attachment rate, while the dashed curve shows the detachment rate as a
function of the droplet radius. At small radii, the detachment
dominates and the droplet tends to evaporate. At large radii, the
attachment dominates and the drop tends to grow in size. There exist a
radius where the rates are equal and the droplet is stable. This is
called the critical radius $r_{\rm{cr}}$, and the droplet, called
critical cluster, contains a critical number of monomers.}
\label{fig:rcr}}
\end{figure}

The rate of detachment of monomers from the condensed phase is much
harder - if at all possible - to compute a-priori. It can be obtained
by comparing the experimentally determined saturation density
$n_{X,\rm{eq}}(T)$ with the theoretical attachment rate from
Eq.~\ref{eq:attach} (see, e.g., Kashchiev 2000, hereinafter K00, and
references therein):
\begin{equation}
\frac{dn_{\rm{out}}}{dt\;dA} = k_s n_{X,\rm{eq}}(T)\sqrt{
\frac{kT}{2\pi m_X}}
\label{eq:detach}
\end{equation}

Equation~\ref{eq:detach}, which is formally identical to
Eq.~\ref{eq:attach}, becomes interesting if we assume that the
detachment rate from the condensed phase does not depend on the
properties of the vapor phase (pressure, temperature, saturation),
This seems to be a good assumption, since the properties and dynamics
of a condensed material should be rathher independent from the
low-density gas surrounding it. However, the detachment rate may
depend slightly on the vapor conditions (pressure, temperature,
density) through their effect on the properties of the condensed
material (density, conductivity, surface tension). We neglect here
these dependencies since there is no established theory to model them.
With this assumption, Eq.~\ref{eq:detach} give us the detachment rate
as a function of the properties of the condensed phase only, as long
as the saturation density as a function of temperature is known either
theoretically or experimentally.

So far we discussed the equilibrium between a flat surface and a
vapor. Phase transitions cannot take place in a single bulk event, but
must be realized through microscopical nucleation of the new phase,
due to energetic constraints (K00).  Equation~\ref{eq:detach} is
modified if the surface that separate the vapor from the condensed
phase is not flat. Let us consider a spherical droplet of radius
$r$. Both thermodynamic (Vehkam\"aki 2006) and kinetic consideration
(K00) show that the saturation density for a droplet is larger than
for a flat surface. In the thermodynamic case, this is due to the
surface energy component, while in the kinetic approach, the effect
can be explained through the fact that the binding energy of a surface
monomer is smaller than for a bulk one, and so it is easier to eject a
monomer for a small droplet than for a flat surface (at the same
temperature).  As long as the number of molecules in the droplet
is large enough to allow for the definition of a surface, it can be
shown that (K00):
\begin{equation}
n_{X,\rm{eq}}(r,T)=n_{X,\rm{eq}}(T)\,e^{\frac{2\sigma{}v_0}{rkT}}
\label{eq:sigma}
\end{equation}
where $\sigma$ is the surface tension of the condensed phase and $v_0$
the volume occupied by one molecule in the condensed phase.  Since the
rate of impacts of vapor molecules on the surface does not depend on
the curvature, and assuming that the sticking coefficient does not
either, we can easily find the detachment rate as a function of
temperature as:
\begin{equation}
\frac{dn_{\rm{out}}(r)}{dt\;dA} = k_s n_{X,\rm{eq}}(T) \sqrt{
\frac{kT}{2\pi m_X}}\,e^{\frac{2\sigma{}v_0}{rkT}}
\label{eq:detachr}
\end{equation}
where we used the so-called capillary approximation, i.e., we assume
that $\sigma$ does not depend on $r$. Even though not explicit, the
ejection rate of Eq.~\ref{eq:detachr} depends strongly on the grain
temperature through the equilibrium density $n_{X,\rm{eq}}(T)$. The
equilibrium density depends exponentially on the temperature, rougly
as $n_{X,\rm{eq}}(T)\propto e^{-A/kT}$, where $A$ is a positive,
material dependent, constant.

Figure~\ref{fig:rcr} compares the attachment and detachment rates of
monomers for water droplets at a temperature $T=293~K$ and at
saturation $S=n_X/n_{X,\rm{eq}}=4$. We adopted $\sigma=78$~dyne/cm,
$v_0=m_{H_2O}/\rho=3\times10^{-23}$~cm$^3$, $k_s=1$, and we obtained
the equilibrium pressure from the CHERIC (Chemical Engineering
Research Information Center) web site\footnote{\tt
http://www.cheric.org/research/kdb/hcprop/cmpsrch.php}.  At small
radii, the detachment rate is larger than the attachment rate and the
droplet will tend to evaporate, while at large radii the droplet tends
to grow. At the critical radius $r_{\rm{cr}}$ the two rates are equal
and the droplet is in equilibrium. The droplet with this radius is
also called {\it critical cluster}. It is easy to see that the problem
of nucleation reduces to the difficulty of creating droplets big
enough to be super-critical, since once the critical radius is passed,
the droplet will grow into the condensed phase.

The fact that a droplet (or nucleus) smaller than $r_{\rm{cr}}$ tends
to evaporate does not mean that it is impossible to nucleate in the
absence of pre-existing super-critical droplets. It exists, in fact,
the possibility that by statistical fluctuations a nucleus grows big
enough to reach the critical radius and subsequently evolve in a
stable droplet. For a super-saturated vapor, it can be shown that the
stationary distribution of nuclei is given by (K00 and references
therein):
\begin{equation}
\frac{dN}{dn}=n_{X}
\frac{\prod_{i=1}^{n-1} f_i}{\prod_{i=2}^{n}g_i}
\left[1+\sum_{i=2}^{\infty}\left(\prod_{j=2}^{i}\frac{g_j}{f_j}
\right)\right]^{-1}\,
\sum_{i=n}^{\infty}\left(\prod_{j=2}^{i}\frac{g_j}{f_j}\right)
\label{eq:distr}
\end{equation}
where $n$ is the number of molecules in the droplet,
\begin{equation}
f_n=4\pi\,r^2\frac{dn_{\rm{in}}}{dt\;dA}
\end{equation}
is the attachment rate for the droplet with $n=(4\pi/3)\,r^3/v_0$
molecules, and
\begin{equation}
g_n=4\pi\,r^2\frac{dn_{\rm{out}}}{dt\;dA}
\end{equation}
is the detachment rate for the same droplet. Equation~\ref{eq:distr}
is quite complex. We refer to K00 for a more detailed discussion. We
here simply notice that since the detachment and attachment rates
enter only through their ratio, the stationary distribution does not
depend on the unknown sticking coefficient $k_s$.

Another fundamental result of the kinetic approach is the nucleation
rate $J$, i.e., the rate at which new nuclei of the condensed phase
are created in the vapor. It reads (K00):
\begin{equation}
J=n_{X}\,f_1\,
\left[1+\sum_{i=2}^{\infty}\left(\prod_{j=2}^{i}\frac{g_j}{f_j}
\right)\right]^{-1}
\label{eq:nucl}
\end{equation}
and it now depends linearly on the sticking coefficient $k_s$ through
$f_1$.  In Eq.~\ref{eq:distr} and~\ref{eq:nucl}, the infinite limits
of summation can be substituted to summation up to twice the number of
molecules in the critical cluster (as long as only the distribution up
to the critical cluster is of interest). Finally, Eq.~\ref{eq:distr}
and~\ref{eq:nucl} are valid in the nucleation stage, i.e., when the
grain growth and destruction proceeds through attachment or detachment
of monomers. In the coalescence stage, when grain-grain collisions are
relevant, different equations must be adopted.

\begin{figure}
\psfig{file=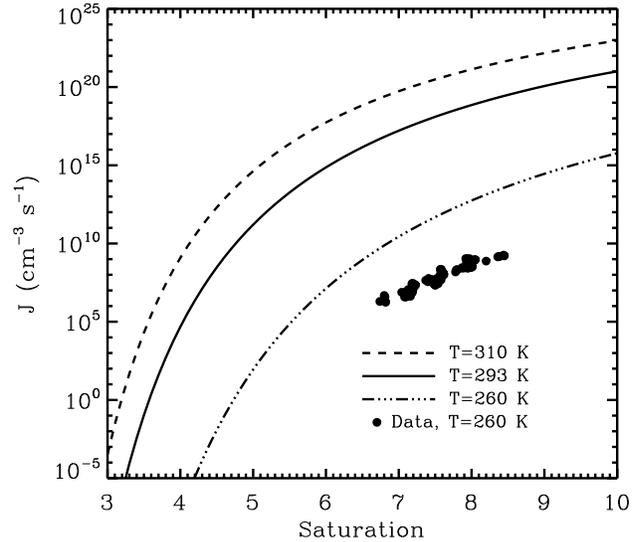,width=\columnwidth}
\caption{{Nucleation rate for water as a function of saturation for
three different temperatures: $T=260$~K (dot-dashed line); $T=293$~K
(solid line), and $T=310$~K (dashed line). Experimental results from
W\"olk \& Stray (2001) for $T=260$~K are shown with solid dots.}
\label{fig:j}}
\end{figure}

Figure~\ref{fig:j} shows the nucleation rate as a function of
saturation $S$ for water droplets in vapor. The radius of the critical
cluster is computed inverting equating the attachment and detachment
rates (Eq.~\ref{eq:attach} and~\ref{eq:detachr}) and reads:
\begin{equation}
r_{\rm{cr}}=\frac{2\sigma v_0}{kT{\rm{ln}}S}
\label{eq:rcr}
\end{equation}
The fact that the nucleation rate increases with saturation is a
direct consequence of the decrease of the critical radius. A smaller
critical radius requires less statistically unfavored random
associations to reach the stable configuration and therefore increases
the nucleation rate. In Eq.~\ref{eq:nucl}, the values of $f_i$ are
directly proportional to saturation, while the $g_i$ do not depend on
$S$. This difference brings about, at the mathematical level, the
dependence of the nucleation rate on saturation. Three temperatures
are shown in Fig.~\ref{fig:j}: $T=293$~K (solid), $T=260$~K (dash-dot)
and $T=310$~K (dashed). The solid dots show the result of a water
nucleation experiment (W\"olk \& Stray 2001) at $T=260$~K. The
comparison of the theory with the data shows that there is a
discrepancy of several orders of magnitude between data and
experiment. This could be due to $k_s\ll1$. However, such an
explanation is unlikely for water droplets (see Batista et
al. 2005). It is rather believed that the discrepancy is due to the
failure of the capillary approximation for droplets with a very small
$n$. A larger surface tension for small droplets is required to make
the theory consistent with observations. When commenting the
quantitative results of the nucleation theory, we shall keep in mind
that reality may differ by few orders of magnitude from the theory.

\section{Thermal balance}

All the theoretical framework discussed in the above section was
developed in an attempt to describe natural and laboratory evidence of
nucleation (e.g. rain). In the astrophysical environment, conditions
can be very different from the laboratory and the theory needs some
adjustments. First we consider that, in general, astrophysical vapors
are immersed in an overwhelming quantity of inert H atoms (or, at
least, inert in the condensation process).  Second, at astrophysical
densities the dust particles are at a lower temperature than the gas
phase.  In this section, we compute the effect of the non thermal
equilibrium between gas and droplets (hereinafter dust grains). There
are several components affecting the grain temperature, the most
important ones are heating by collisions with gas phase particles and
heating/cooling through the interaction with the radiation field.

We here consider two effects: heating by collisions with gas particles
and cooling due to the emission of radiation. We neglect radiation
heating, i.e. we assume the radiation field is diluted and unimportant
(radiation is a sink of energy). The cooling due to radiation is given
by:
\begin{equation}
\frac{dE}{dt} = 4\pi r^2 \sigma_{\rm{SB}} \, {\cal A}(r,T_s) \, T_s^4
\label{eq:eout}
\end{equation}
where $\sigma_{\rm{SB}}=5.67\times10^{-5}$~erg~cm$^{-2}$ K$^{-4}$
s$^{-1}$ is the Stefan-Boltzmann constant, ${\cal A}\le1$ a factor
that takes into account that the grain does not emit as a black body
at all frequencies and $T_s$ is the temperature of the grain. ${\cal
A}(r,T)$ is given by (Laor \& Draine 1993):
\begin{equation}
{\cal A}(r,T) = \frac{\int Q_{\rm{abs}}(\lambda) B_\lambda(T)
\,d\lambda}{\int B_\lambda(T)\,d\lambda}
\label{eq:cala}
\end{equation}
where $Q_{\rm{abs}}(\lambda)$ is the absorption efficiency and
$B_\lambda(T)$ is the black-body spectrum.

\begin{figure}
\psfig{file=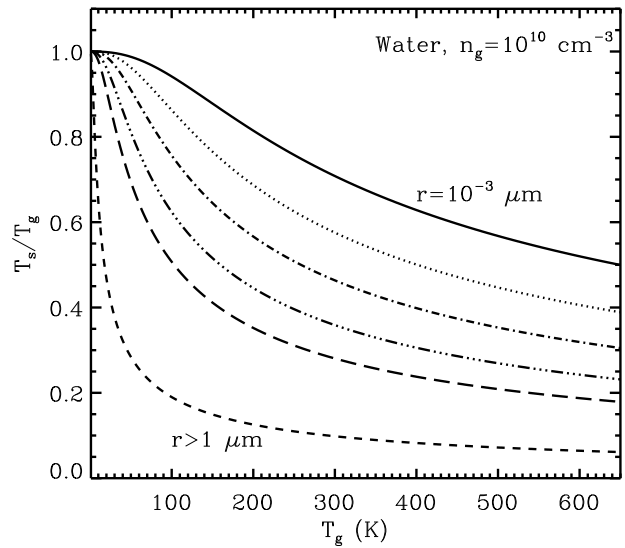,width=\columnwidth}
\caption{{Temperature of water droplets in a vapor as a function of
the vapor temperature $T_g$. The temperature depends on the droplet
radius since the radiation emitted is proportional to the droplet
opacity, which scales linearly with the radius. From top to bottom,
the lines show droplets of radius: $10^{-3}$, $3\times10^{-3}$,
$10^{-2}$, $3\times10^{-2}$, $10^{-1}$ and $\ge1\mu$m. These radii
correspond to 3,10,30,100,300, and 3000 inter-molecular distances,
respectively. 
}
\label{fig:tgts}}
\end{figure}

To compute the heating of the grain due to collisions, we consider the
astrophysical scenario, where most of the collision between the grain
and the gas particles do not lead to attachment because of the
overwhelming abundance of H over the grain monomers. We also neglect
vibrational and rotational energy of the gas particles (again, we
suppose to be dominated by H).  Finally, we neglect the sputtering of
monomers from the cluster that can result from the collision with a
high speed gas particle. Such effect is known to be relevant in shock
destruction of dust particles but is usually neglected in the
classical theory of nucleation (K00). A detailed treatment of the
effect requires a detailed description of the energy levels of the
clusters, a formidable task that is beyond the scope of this paper.
We obtain:
\begin{eqnarray}
\frac{dE}{dt}&=&4\pi r^2\int_0^\infty
\frac{n_{\rm{g}}v}{4}p(v)\left(\frac{1}{2}\mu m_Hv^2-\frac{3}
{2}kT_{\rm{s}}\right)\,dv  \\
&=& 4\pi\,r^2\sqrt{\frac{2kT_{\rm{g}}}{\pi \mu m_H}}n_{\rm{g}}kT_g
\left(\frac{T_{\rm{g}}-T_{\rm{s}}}{T_{\rm{g}}}\right)
\label{eq:ein}
\end{eqnarray}
where $p(v)$ is the Maxwellian distribution of velocities, $m_H$ is
the hydrogen mass, and $\mu\sim1.2$ the mean atomic weight of the
gas. The subscript $_s$ always refer to the solid (grain) phase, while
the suffix $_g$ always refers to the gas (vapor) phase.

The balance between losses and gains of energy gives us an implicit
equation for the temperature of the grain as a function of the gas
pressure and temperature. Assuming a single equilibrium temperature
for small grains is an oversimplification (Guhathakurta \& Draine
1989). A broad range of temperatures would likely increase the
nucleation rate since the cloder-than-average grains could grow into
stable grains even for conditions where the average temperature would
not allow for nucleation. On the other hand, hotter-than-average
grains would be destroyed, providing a balancing effect. Neglecting
the temperature distribution we obtain:
\begin{equation}
{\cal A}(r,T_s)\,T_s^4 = \left(\frac{2kT_g}{\pi \mu m_H}\right)^{1/2}
\frac{n_gkT_g}{\sigma_{\rm{SB}}} \frac{T_g-T_s}{T_g}
\label{eq:ts}
\end{equation}
%
%
%
%

\begin{figure}
\psfig{file=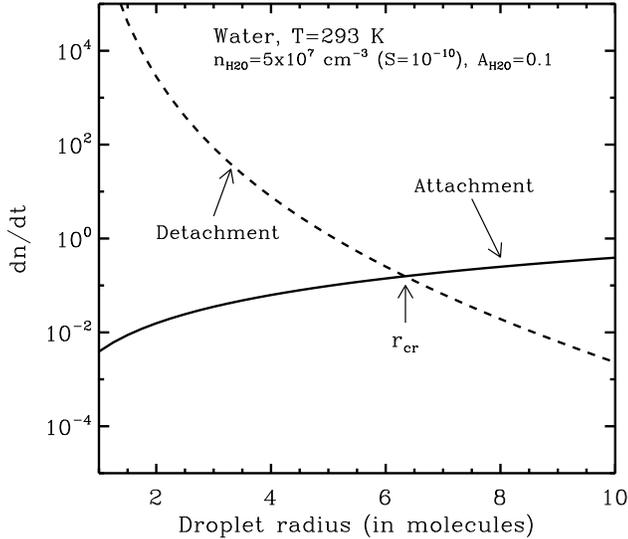,width=\columnwidth}
\caption{{Same as Fig.~\ref{fig:rcr} but for an sub-saturated
vapor. Nucleation is still possible due to the cooling of the
droplet. A critical radius can be found in this case, analogously to
the classical nucleation scenario.}
\label{fig:vart}}
\end{figure}

Figure~\ref{fig:tgts} shows the magnitude of the effect 
for water droplet in a hydrogen dominated gas of total density
$n_g=10^{10}$~cm$^{-3}$. For the largest size we used the
approximation ${\cal A}=1$, while for small water droplets we adopted
the approximation:
\begin{equation}
{\cal A}(r,T) \simeq 1-e^{r\tau_r}\simeq r\tau_r; \qquad r\ll\tau_r^{-1}
\end{equation}
where $\tau_r$ is the opacity per unit length in the medium. For pure
water we have ${\cal A}(r) \simeq 1200\,r$ in the temperature range
$250<T<600$~K (pure water absorption coefficients are taken from the
Oregon Medical Laser Center-OMLC web page\footnote{\label{pro}\tt
http://omlc.ogi.edu/spectra/water/}).

\section{Kinetic equilibrium}

As noted by Draine (1981), the departure from thermal equilibrium can
have important effects on the kinetic balance of the clusters.  Even
though the effect of the grain radiative cooling is not exceedingly
large (see Fig.~\ref{fig:tgts}), especially for the small grains, the
detachment rates change exponentially with temperature. This implies
that even a small temperature change can make the difference between a
nucleating mixture and a non-nucleating one. An additional
complication is caused by the fact that while the grain temperature
depends on the total gas density $n_g$, the attachment rate depends on
the partial gas density of the compound of interest $n_X=\Xi_X\,n_g$,
where we have defined the number abundance of the compound $\Xi_X$.

Figure~\ref{fig:vart} shows the effect of the temperature change in
the dynamic equilibrium for water droplets in a vapor with temperature
$T_g=293$~K, $n_g=5\times10^8$~cm$^{-3}$, and $\Xi_{H_2O}=0.1$. In
such conditions, the saturation\footnote{The saturation is defined as
$S=n_X/n_{X,{\rm{eq}}}(T_g)$ also in non-LTE conditions, but $S=1$
loses in this case the meaning of threshold for the nucleation
process.} is $S=10^{-10}$ and the vapor should not nucleate. Indeed,
for $T_s=T_g$, the detachment rate would be $\sim10^{10}$, about ten
orders of magnitude larger than the attachment rate, and droplets
would evaporate. Due to the approximately exponential dependence of
the experimental saturation density with temperature, the change in
temperature of the droplets suppresses the detachment rate. The
situation appears now qualitatively similar to the nucleating
condition in Fig.~\ref{fig:rcr}. At small radii, the detachment rate
is larger than the attachment rate, and the droplets evaporate. A
critical radius $r_{\rm{cr}}$ does exist, where the two rates are
identical and the droplet is in equilibrium. The critical radius
is now given by a modified version of Eq.~\ref{eq:rcr}, since the gas
and grain temperatures are different:
\begin{equation}
r_{\rm{cr}} = \frac{2\sigma v_0}{kT_s\ln Z}
\end{equation}
where
\begin{equation}
Z=\frac{n_X}{n_{X,{\rm{eq}}}(T_s)}\sqrt{\frac{T_g}{T_s}}
\end{equation}

In the same way as in the LTE case, droplets larger than the critical
radius grow due to the larger attachment rate with respect to the
detachment rate.

Figure~\ref{fig:vart} was computed solving numerically Eq.~\ref{eq:ts}
for the temperature with ${\cal A}(r,T)$ obtained from
Eq.~\ref{eq:cala} with water data taken from the OMLC web page (see
footnote~\ref{pro}).  Again, $k_s=1$ was assumed. Relaxing this
assumption would change the normalization of the two curves so that
$r_{\rm{cr}}$ would remain constant.

\section{Nucleation}

\begin{figure}
\psfig{file=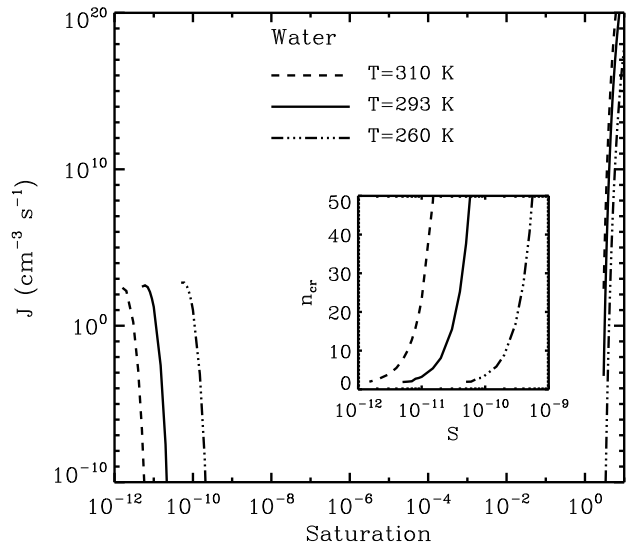,width=\columnwidth}
\caption{{Same as Fig.~\ref{fig:j} but extended to very low
saturation. The nucleation rate is non-negligible and nucleation is
possible. The inset shows the number of monomers in the critical
cluster as a function of saturation. Contrary to the classical case,
the critical radius increase with the saturation, in this case. The
curves on the right side of the plot are the results for LTE
nucleation from Fig.~\ref{fig:j}.}
\label{fig:j2}}
\end{figure}

The nucleation rate for the sub-saturated nucleating vapor can be
computed using Eq.~\ref{eq:nucl} given the attachment rate for the gas
temperature and the detachment rate computed at the grain
temperature. Figure~\ref{fig:j2} shows an extension of
Fig.~\ref{fig:j} to much lower saturations (and therefore
densities). When saturation drops below unity, nucleation does not
take place, until at very low densities and saturations non-LTE
nucleation sets in. The nucleation rate is somewhat small, since the
attachment rate is low at low saturation, but the detachment rate is
suppressed by the decreased temperature of the grains and therefore
nucleation is possible. It must be emphasized that the theory becomes
more and more inaccurate as we approach the limit of a small number of
monomers in the critical cluster. In this case, many of the
approximations made to derive the theory lose validity. First, the
opacity per unit length looses meaning, and the cooling rate is not
computed accurately. Second, the cluster does not have a well-defined
surface any more and the capillary approximation for the surface
tension breaks down. For this reason we did not include in
Fig.~\ref{fig:j2} the very low saturation limit, since the number of
monomers in the critical cluster becomes close to unity. Nucleation is
still possible there, but computing the rate in the atomistic
approximation is beyond the goal of this paper and, for what matters,
a very controversial issue.  We show in the inset of Fig.~\ref{fig:j2}
the number of monomers in the critical cluster as a function of
saturation in the non-LTE nucleating region. Calculation have been
stopped at the limit $n_{\rm{cr}}=2$, but for $n_{\rm{cr}}\lsim5$ the
results should be taken with caution.

\begin{figure}
\psfig{file=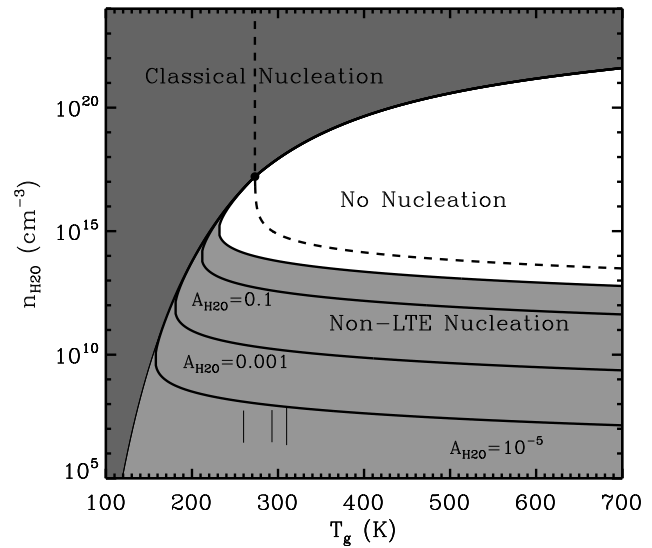,width=\columnwidth}
\caption{{Nucleation phase diagram for water droplets. The classical
vapor pressure line is shown as a thin solid line. The new
equilibrium is shown as a thick solid line. Only the region inside the
line does not produce any nucleation. The three thin vertical lines in
the lower left corner show the values for which the nucleation rate is
shown in Fig.~\ref{fig:j2}. The dashed line shows the ice formation
boundary: only the droplets in the upper right corner are liquid
(assuming that water freezes at 273.15 K independently of pressure).}
\label{fig:diag}}
\end{figure}

Finally, we show in Fig.~\ref{fig:diag} the new conditions for water
droplet nucleation in the $n-T$ plane. As explained above, nucleation
in the non-LTE area depends on the abundances $\Xi_{H_2O}$.  In dark
gray we show the region where nucleation is possible in the classical
view. In light gray, we show the area where non-LTE nucleation is
possible in the limit $\Xi_{H_2O}\to1$. Solid lines show how the area
is modified in the cases of $\Xi_{H_2O}=10^{-1}$, $10^{-3}$, and
$10^{-5}$. The three thin vertical lines in the lower left part of the
diagram show the conditions under which nucleation rates have been
computed in Fig.~\ref{fig:j2}. It is important to note that even
though nucleation is possible at higher densities, the nucleation rate
becomes non-negligible only for densities roughly equal or smaller
than $n_{H_2O}=10^{8}$~cm$^{-3}$. The dependence of nucleation on
density, for a given temperature, is quite complex. At very high
densities, the conditions of temperature equilibrium are maintained by
the high rate of collisional heating and classical LTE nucleation
takes place. As the density is decreased, two different things can
happen. At high temperature (the region in Fig.~\ref{fig:diag}
and~\ref{fig:c} where a white area is present), the vapor becomes
sub-saturated and nucleation is halted. Decreasing the density even
further allows for the cooling of the sub-critical grains, and
nucleation in non-LTE sets in. If the temperature is lower, non-LTE
conditions apply even in saturated vapors. In that case, the vapor is
always nucleating (see Fig.~\ref{fig:jj}).  The dashed line in
Fig.~\ref{fig:diag} shows the region where the temperature of the
droplets is equal to the freezing point of water (assuming it freezes
at $0^\circ$~C independent on pressure). Non-LTE nucleation results
therefore in ice and not liquid droplets.  Finally, the above
treatment is valid only if there are no pre-existing grains
(e.g. graphite or silicates) that can work as seeds for heterogeneous
nucleation. In that case, the presence of already cold grains could
result in the growth of ice mantels rather than in the nucleation of
new droplets.

\subsection{Graphite}

Even though water has an astrophysical relevance for dust, especially
in covering refractory grains with ice mantles, we now consider a more
pregnant example: graphite grains condensing from carbon atoms in the
gas phase. For simplicity, we consider a gas made only of hydrogen
with solar carbon abundance: $\Xi_C=3.3\times10^{-4}$ (Grevesse \&
Sauval 1998). We use the vapor pressure measurement from Brewer,
Gilles \& Jenkins (1948) and the surface tension
$\sigma_C=34.6$~dyne/cm (Morcos 1972).

\begin{figure}
\psfig{file=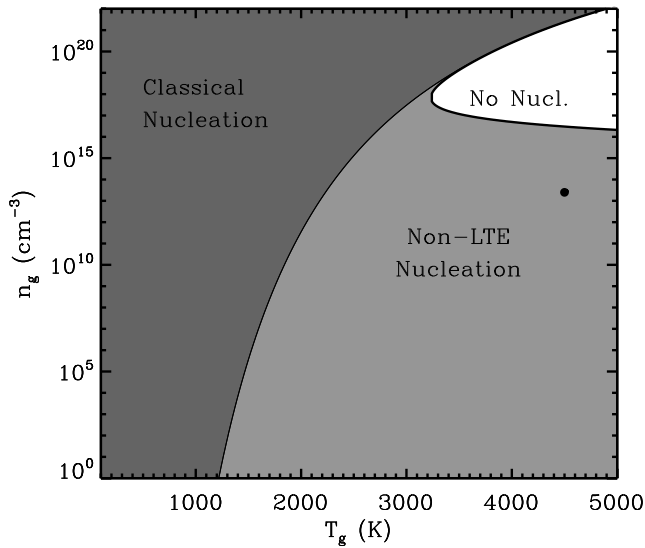,width=\columnwidth}
\caption{{Nucleation phase diagram for graphite grain in a hydrogen
gas with carbon number abundance $\Xi_C=3.3\times10^{-4}$. As for the
case of water analyzed above, there exist a big region at high
temperatures and low densities where graphite grains can nucleate in
non-LTE conditions with the gas. The point marks the conditions for
which the nucleation rate $J\approx7\times10^4$~cm$^{-3}$~s$^{-1}$ has
been computed in the text.}
\label{fig:c}}
\end{figure}

Figure~\ref{fig:c} is analogous to Fig.~\ref{fig:diag}, but shows the
phase diagram of graphite. Note also that the y-axis reports the total
density of the gas and not the partial density of carbon
atoms. Analogously to the case of water, we see that graphite grains
can form in non-LTE at temperatures much larger than in the classical
scenario.

To compute nucleation rates in non-LTE, we need to derive the factor
${\cal A}$ for graphite. We use the tables\footnote{\tt
http://www.astro.princeton.edu/$\sim$draine/dust/dust.diel.html} of
$Q_{\rm{abs}}$ from Draine \& Lee (1984) and Laor \& Draine (1993). In
some cases, however, the grain sizes given in the tables are not small
enough (the smallest grain considered in the tables has $r=0.001\mu$m
while the smallest we consider has $r=0.00013\mu$m). In that case we
extrapolate to smaller radii in the optically thin limit assuming
$Q_{\rm{abs}}(\lambda,r)=1-\exp(-r\tau_r)$.

In the case of graphite, nucleation in the non-LTE region can be very
efficient. For a gas density $n_g=2.5\times10^{13}$~cm$^{-3}$ and
temperature $T_g=4500$~K (see solid dot in Fig.~\ref{fig:c}), the
critical cluster has $\approx7$ carbon atoms and the nucleation rate
is $J\approx7\times10^4$~cm$^{-3}$~s$^{-1}$.

\begin{figure}
\psfig{file=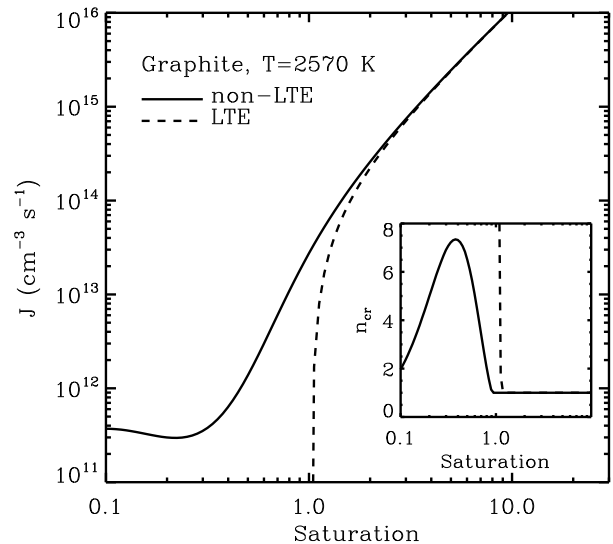,width=\columnwidth}
\caption{{Nucleation rate in a carbon-hydrogen mixture with solar
carbon abundance at $T=2570$~K. The dashed line shows the classical
result for ``true equilibrium'' nucleation, while the solid line shows
the result of non-LTE nucleation. Even for $S\gsim1$, where LTE
nucleation takes place, allowing for the grain cooling increases the
nucleation rate.  The inset shows the number of monomers in the
critical cluster for the two cases as a function fo the
saturation. Not that computations in the super-saturated regime suffer
the uncertainties of the few monomers limit for both LTE and non-LTE
assumptions.}
\label{fig:jj}}
\end{figure}

Another important aspect of Fig.~\ref{fig:c} is that there is a vast
temperature range where there is no non-nucleating region between the
classical and non-LTE nucleating regions. This was true for water as
well, but in a narrower region. For $T_g<3250$~K in the
carbon-hydrogen mixture considered, the grain temperature is smaller
than the gas temperature even in the classical nucleating region. This
implies that the nucleation rate is different from the classical value
and the cooling of the grains has to be considered even in the
classical region. Figure~\ref{fig:jj} shows the nucleation rate for
$T=2500$~K in the saturation range $0.1\le S\le10$. Even though a
mixture in true equilibrium does nucleate for $S>1$, non-LTE effects
increase the rate.

\section{Discussion and conclusions}

We have considered the effect of grain (cluster, droplet) cooling in
the nucleation of liquid and solid phases in vapors. We find that the
effect can be dramatic on the nucleation rate and on the nucleation
phase diagram, allowing for nucleation in large regions of the
parameter space that are classically considered to be non-nucleating.

As is in general true for nucleation, there are several limits and
approximations that we should bear in mind when considering the theory
from the quantitative point of view. As exemplified by the comparison
of the data with the theory in Fig.~\ref{fig:j}, several orders of
magnitude can separate the nucleation rate prediction from the
observations. The controversial points are:

\begin{itemize}
\item {\it Sticking coefficients} --- Most of the figures and
computations in this work assume $k_s=1$. This is not always true
(Batista et al. 2005). In addition to its dependence on temperature,
the sticking coefficient may depend on the size of the cluster. A big
cluster could more easily absorb the extra kinetic energy of the
incoming monomer, compared to a small cluster (K00), and therefore
$k_s$ may be significantly smaller than unity for very small clusters.
\item {\it Capillary approximation} --- The capillary approximation,
i.e., the assumption that the surface tension does not depend on the
cluster size, is very controversial, and a change in the surface
energy for very small clusters could result in big changes on the
nucleation rate. For example, the discrepancy in Fig.~\ref{fig:j}
could be solved by assuming a larger surface tension for the very
small water droplet. In addition, macromolecules do not even have a
properly defined surface, and the whole concept does not
apply. Finally, the exponential factor in Eq.~\ref{eq:sigma} depends
on the assumption that the clusters are spherical. A different ratio
of the surface to the volume would modify this term. This is likely
for small graphite clusters, since graphite tends to aggregate in a
planar form.
\item {\it Detachment rate for small clusters} --- In this paper, and in
most nucleation theory, the detachment rate is computed by propagating
to very small clusters the detachment rate of macroscopic bodies. It
is very likely that, in the very small limit, the detachment is
governed by completely different processes. Let us analyze the two
limits. For a macroscopic body, the number of monomers is so large
that monomers with a statistically higher energy can detach since
their energy is larger than the binding energy. In the opposite limit
of a dimer, the detachment has to be due to an external action: either
a collision with a fast monomer or with a photon or with another
cluster. In this limit destructive collisions have to be considered.
\item {\it Cooling of macromolecules} --- This is a new problem that
arises when the cooling of the grains is considered. We have assumed
that the grains cool as modified black bodies down to the smallest
sizes. When the number of monomers in the cluster becomes small, the
cooling will not be through a continuum spectrum, but through lines
and bands. A more refined treatment of cooling is necessary to compute
accurate nucleation rates.
\item It should finally be kept in mind that we allowed for the
complete cooling of the grains, neglecting the effects of a background
radiation field in setting a lower limit to the temperature. In
addition, we neglected the fact that the temperature of small grains
is largely stochastic and that at high temperature some collisions
between grains and gas particles can result in sputtering rather than
accretion.
\end{itemize}

Despite all these caveats, the main result of this paper holds: the
region where a vapor spontaneously nucleate is not limited to the
classic region, when nucleation takes place in thermal equilibrium and
$T_g=T_s$. Allowing for the clusters to cool inhibits the detachment
of monomers from the clusters and allow for nucleation at higher
temperatures and lower densities than in the classical scenario. The
nucleation rates tend to be orders of magnitude smaller than those
derived in thermal equilibrium, but provide a non-negligible
correction to the LTE rate for mildly super-saturated vapors with
$S\gsim1$, especially at low temperature.

In the astrophysical scenario small nucleation rates are not a big
worry. The ISM of our Galaxy contains approximately 1 per cent of its
mass in dust grains (Mathis et al. 1977). This corresponds to
approximately one dust particle every cubic meter (assuming a
power-law grain size distribution as in Mathis et al. 1977). However,
in the present day Universe the ISM is already polluted with dust and
in the presence of seed grains it is likely that the process of mantle
growth dominates over non-LTE nucleation in the diffuse medium. The
process of non-LTE nucleation may therefore be important at high
redshift, when the ISM is first polluted with metals by supernova
explosions. It is unclear if supernov\ae\ do generate dust by
themselves, and even more whether the generated dust can survive the
sputtering in the forward-reverse shock systems (Bianchi \& Schneider
2007; Nath, Laskar \& Shull 2007). In the case that supernov\ae\ do
mainly pollute the ISM with metals but with no or a negligible
quantity of dust, non-LTE nucleation could become the dominant process
of dust nucleation in the early universe. Detailed estimates of
nucleation in the various scenarios require a more detailed
understanding of the properties of the very small nuclei and are
beyond the scope of this paper.

\section*{Acknowledgements}
I thank the anonymous referee for a thorough review and constructive
comments that improved the contents of this paper. I am grateful to
Mike Shull, Nahum Arav and Bruce Draine for useful suggestions and
discussions.


\begin{thebibliography}{99}

\bibitem{} Abraham F. F., 1974, Homogeneous nucleation theory,
  Academic Press, New York

\bibitem{} Batista E. R., Ayotte P., Bili\'c A., Kay B. D., J\'onsson
  H, 2005, PRL, 95, 223201

\bibitem{} Becker R., D\"oring W., 1935, Ann. Physik, 24, 719

\bibitem{} Bianchi S., Schneider R., 2007, MNRAS, 378, 973

\bibitem{} Brewer L., Gilles P. W., Jenkins F. A., 1948,
  J. Chem. Phys., 16, 797

\bibitem{} Cardelli J.~A., Clayton G.~C., Mathis J.~S., 1989, ApJ,
  345, 245

\bibitem{} Draine B.~T., 1979, Ap\&SS, 65, 313

\bibitem{} Draine B.~T., 1981, ASSL, 88, 317

\bibitem{} Draine B. T., Lee H. M., 1984, ApJ, 285, 89

\bibitem{} Draine B.~T., 1989, eidr.proc, 103

\bibitem{} Draine B.~T., 2003, ARA\&A, 41, 241

\bibitem{} Edminds M. G., 2001, MNRAS, 328, 223

\bibitem{} Elvis M., Marengo M., Karovska M., 2002, ApJ, 567, L107

\bibitem{} Feder J, Russell K. C., Lothe J., Pound G. M., 1966,
Advances in Physics, 15:57, 111

\bibitem{} Gail H.-P., Sedlmayr E., 1986, A\&A, 166, 225

\bibitem{} Grevesse N., Sauval A. J., 1998, Space Science Reviews, 85,
  161

\bibitem{} Kashkiev D., 2000, ``Nucleation: basic theory with
  applications'', Butterworth-Heinemann, Oxford

\bibitem{} Kozasa T., Hasegawa H., Nomoto K., 1989, ApJ, 344, 325

\bibitem{} Kozasa T., Hasegawa H., Nomoto K., 1991, A\&A, 249, 474

\bibitem{} Laor A., Draine B. T., 1993, ApJ, 402, 441

\bibitem{} Mathis J.~S., Rumpl W., Nordsieck K.~H., 1977, ApJ, 217,
  425

\bibitem{} Mathis J.~S., 1990, ARA\&A, 28, 37

\bibitem{} McKee C., 1989, IAUS, 135, 431

\bibitem{} Morcos I., 1972, J. Chem. Phys., 57, 1801

\bibitem{} Nath B. B., Laskar T., Shull J. M., 2007, ApJ submitted

\bibitem{} Nowakowski B., Ruckenstein E., 1991a, J. Chem. Phys., 94,
  1397

\bibitem{} Nowakowski B., Ruckenstein E., 1991b, J. Chem. Phys., 94,
  8487

\bibitem{} Nozawa T., Kozasa T., Umeda H., Maeda K., Nomoto K., 2003,
  ApJ, 598, 785

\bibitem{} Salpeter E.~E., 1977, ARA\&A, 15, 267

\bibitem{} Schneider R., Ferrara A., 
Salvaterra R., 2004, MNRAS,
  351, 1379

\bibitem{} Stein W.~A., Soifer B.~T., 1983, ARA\&A, 21, 177

\bibitem{} Todini P., Ferrara A., 2001, MNRAS, 325, 726

\bibitem{} Vehkam\"aki H. 2006, ``Classical nucleation theory in
  multicomponent systems'', Springer, Heidelberg

\bibitem{} Weingartner J.~C., Draine B.~T., 2001, ApJ, 548, 296

\bibitem{} W\"olk J., Strey R., 2001, J. Phys. Chem. B, 105, 11683

\end{thebibliography}
\end{document}